\documentclass[12pt]{article}
\usepackage[table]{xcolor}
\usepackage{amsthm,amsfonts}
\usepackage{cite}
\usepackage{avant}
\usepackage{amssymb}
\usepackage{multirow,booktabs}
\usepackage{caption}
\usepackage{subcaption}

\usepackage{pifont}\usepackage{amsmath}
\usepackage[mathscr]{eucal}
\usepackage[all]{xy}
\usepackage{listing}
\usepackage{pdfpages}
\xyoption{arc}
\xyoption{curve}
\numberwithin{equation}{section}
		\newfont{\spfnt}{punk12}
		\newcommand\email[4]{#1@#2.#3.#4}

		\renewcommand\hat{\widehat}

		\newfont\sheafnt{rsfs10}

\begin{document}
			\title{Fermionic spectrum from a domain wall in five dimensions}
			\author{
				Subir Mukhopadhyay\thanks{\email{subirkm}{gmail}{com}{}} and Nishal Rai
				\thanks{\email{nishalrai10}{gmail}{com}{}} \\
				\small Department of Physics, Sikkim University, 6th Mile, Gangtok 737102
			}
			\date{}
			\maketitle
			\begin{abstract}
				\small
				\noindent
				We study fermions in a domain wall backgrounds  in five dimensional supergravity, which is similar to zero temperature limit of holographic superconductor. We find the fermionic operators for small charges in the dual four dimensional theory have gapped spectrum.
			\end{abstract}
			\thispagestyle{empty}
			\clearpage

	
		\section{Introduction}

Holographic methods \cite{Maldacena:1997re,Gubser:1998bc,Witten:1998qj} are very useful tools for studying strongly coupled fermionic systems. These have been effectively used to study Fermi surfaces \cite{Lee:2008xf,Cubrovic:2009ye,Liu:2009dm} with black hole background on the gravity side, leading to appearance of  holographic Fermi and  non Fermi liquids with different scaling behaviors of excitations. Various aspects of the non-Fermi liquids were studied \cite{Faulkner:2009wj,Edalati:2010ge,Edalati:2010ww}, with effects of variation of different parameters. The approach employed in these works is bottom up, where on the gravity side one considers a custom gravity theory reflecting the appropriate symmetry of the operators in the low energy effective theory. In this approach gravity theory is tailored to produce desired dynamics and advantage lies in its flexibility.

Another approach is top down, where one considers a known string or supergravity model and the advantage is the dual field theory is known. In this  approach, cases of probe branes and  N=2 supergravity theories were studied in \cite{Ammon:2010pg,Jensen:2011su,Gauntlett:2011mf,Belliard:2011qq,Gauntlett:2011wm}. Subsequently, analyses of maximally gauged supergravity theories appeared in literature \cite{DeWolfe:2011aa,DeWolfe:2012uv,DeWolfe:2014ifa,self} at zero temperature leading to Fermi surfaces in the dual theories. These were further extended to computation of Greens function at finite temperatures, giving rise to ungapped spectrum\cite{Cosnier-Horeau:2014qya,self1}. These studies considered backgrounds  having finite entropy at zero temperature.  Later, a model having vanishing entropy at zero temperature was analysed in  \cite{DeWolfe:2013uba} where they found fermionic fluctuations are stable within a gap around Fermi surface. Gapped spectra were also found from the analysis of Green's function at finite temperature for Lifshitz geometry in bottom-up approach\cite{Wu:2013oea,Wu:2014rqa}.
  Discussions of Fermi surfaces in similar context appeared in  \cite{Berkooz:2012qh,Berkooz:2006wc,Berkooz:2008gc}.

A different class of backgrounds were considered in this vein, where symmetry gets broken due to condensation of a charged scalar in the gravity theory. The zero temperature limits of these backgrounds are expected to be domain wall solutions of the supergravity theory \cite{Gubser:2008wz,Gubser:2009cg,Horowitz:2009ij}. Such backgrounds appear in the studies of condensed phase of holographic superconductors and may be related to the non-Fermi liquids. Analyses of spectral function of fermions at zero temperature of a holographic superconductor with condensed scalar appeared in \cite{Chen:2009pt} in bottom up approach and they reported peak-dip-hump structure as found in APRES experiment. \cite{Faulkner:2009am} considered Majorana fermions coupled to itself, as well as to a cooper pair scalar of twice charge and obtained a gapped spectrum. In view of that, it was natural to study whether holographic superconductors constructed from string and M theory \cite{Gubser:2009qm,Gubser:2009gp,Ammon:2010pg,Gauntlett:2009dn,Gauntlett:2009bh} show similar gaps for fermionic spectra \cite{Gubser:2009dt,DeWolfe:2015kma,DeWolfe:2016rxk,Mukhopadhyay:2018gmf}. Behaviour of generic fermions  in the background of a domain wall in four dimension, obtained from compactification of M theory was  studied in \cite{Gubser:2009dt}, giving rise to bands of normalisable modes in the region of space-like momentum.  Analysis of domain wall backgrounds in four dimensional gauged supergravity, dual to Aharony-Bergman-Jafferis-Maldacena (ABJM) theory with a symmetry breaking source appeared in\cite{DeWolfe:2015kma}, leading to both gapped and gapless bands of stable quasi-particles. Similar domain wall solutions, dual to states in ABJM theory with broken $U(1)$ symmetry were studied in \cite{DeWolfe:2016rxk}, where the gap in the spectrum has been attributed to small fermionic charge and interaction between particles and holes.



In the present work we consider a domain wall solution in five dimensional gravity theory given in \cite{Gubser:2009gp}. This theory can be obtained by compactification of type IIB supergravity on a five dimensional squashed Sasaki-Einstein manifold \cite{Gubser:2009qm,Bah:2010cu} after making suitable truncations.  Solution of equation of motion of this truncated theory can be expected to remain a solution when uplifted to the full theory. The domain wall interpolates between two AdS geometries with spontaneously breakdown of a $U(1)$ symmetry and in that respect,  it may corresponds to zero temperature of holographic superconductor. In this background, we consider dynamics of certain fermionic modes that appear in this truncated five dimensional theory. It turns out \cite{Bah:2010cu} that after suitable truncation, the fermionic modes can be separated in different sectors of which there is one with a single fermion, which does not couple to other fermions or gravitini. Using holographic method we have studied spectra of the dual operator. This domain wall solution corresponds to some state in the dual field theory and so it can shed light on the behaviour of fermionic operators there. In addition, this five dimensional theory demonstrates a different kind of couplings between fermions and charged scalars which may have some phenomenological interest \cite{Bah:2010cu}.  In order to keep our study flexible we analyse fermions with different values of charges. From the analysis in the space like region we find for small charge there is no normal mode around $\omega=0$.  As the charge increases normal modes start appearing for $\omega=0$. We have also studied behaviour of the gap with variation of Pauli term. We have studied the spectral function in the time like region and find excitations having a dispersion relation, which is different from that in the space like region. 

 The plan of the article is as follows. In the next section, we briefly describe the domain wall solution that we use as the background. In section 3 and 4 we present Green's function and its numerical computation for different charges respectively. We conclude with a discussion in section 5.

			\section{Domain Wall solution }

In this section we review the domain wall solution found in\cite{Gubser:2009gp}. We consider compactification of type IIB string theory on a product of an anti-de Sitter space and a Sasaki-Einstein manifold, $AdS_5 \times Y$. A consistent truncation gives rise to a five-dimensional theory with bosonic content consisting of metric, a $U(1)$ gauge field and a complex scalar. The action is given by\cite{Gubser:2009gp}
\begin{equation}\begin{split} \label{action}
S = &\frac{1}{2 \kappa_5^2} \int ~ d^5x \sqrt{-g} (  R - \frac{1}{4} F_{\mu\nu}F^{\mu\nu} \\
&- \frac{1}{2} [(\partial_\mu\eta)^2 + \sinh^2\eta (\partial_\mu\theta - \frac{\sqrt{3}}{L} A_\mu)^2] + \frac{3}{L^2} \cosh^2\frac{\eta}{2} (5 - \cosh\eta) 
\end{split}
\end{equation} 
where the complex scalar field is $\eta e^{i\theta}$ and there is also an additional Chern-Simons term. 
The potential $V(\eta) = - \frac{3}{L^2} \cosh^2 \frac{\eta}{2} (5 - \cosh\eta)$  has two extrema,  $\eta = 0$  and $\eta= Log(2 + \sqrt{3})$

In order to obtain domain wall solution consider following ansatz for the metric, gauge field and scalar field,  
\begin{equation} \label{ansatz}
ds^2 = e^{2A} [ - h dt^2 + d{\bf x}^2 ] + \frac{dr^2}{h},
\quad
A = A_t dt,
\quad
\theta = 0.
\end{equation}

The equations of motion following from the action (\ref{action}) and ansatz (\ref{ansatz}) are given by
\begin{equation}
\begin{split}\label{eom}
3 h A^{\prime\prime} &= - \frac{3}{2L^2} \frac{e^{-2A}}{h} \sinh^2\eta A_t^2 - \frac{1}{2} h (\eta^\prime)^2, \\ 
h^{\prime\prime} + 4 A^\prime h^\prime &= e^{-2A} (A_t^\prime )^2 + 2 \frac{3}{2L^2} \frac{e^{-2A}}{h} \sinh^2\eta A_t^2,\\
\eta^{\prime\prime} + ( 4 A^\prime + \frac{h^\prime}{h} ) \eta^\prime &= - 2 \frac{3}{2L^2} \frac{e^{-2A}}{h^2} A_t^2 \sinh\eta \cosh\eta + \frac{1}{h} V^\prime (\eta),\\
A_t^{\prime\prime} + 2 A^\prime A_t^\prime &= \frac{3}{L^2 h} \sinh^2 \eta A_t,\\
\frac{3}{2} [ 4 h (A^\prime)^2 + A^\prime h^\prime ] &= \frac{1}{2h} [ \frac{1}{2} h^2 (\eta^\prime)^2 + \frac{3}{2L^2} e^{-2A} \sinh^2 \eta A_t^2 - \frac{1}{2} e^{-2A} h (A_t^\prime)^2 - h V(\eta) ],
\end{split}
\end{equation}
The last equation is a constraint and if all other equations are satisfied it holds for all values of $r$ provided it is satisfied at some value of $r$ \cite{Gubser:2008wz}.

The domain wall solution interpolates between two extrema of scalar potential, $\eta = 0$ at UV and $\eta= Log(2 + \sqrt{3})$ at IR. The boundary considtions are chosen as follows. At both the extremes the geometries are $AdS_5$ with radii of curvature $L$ and $L_{IR}= 2\sqrt{2}\frac{ L}{3}$ respectively
At IR, $A_t$ vanishes and $A \sim r/L_{IR}$, $h \sim 1$. The infra-red asymptotic behaviour of gauge field and scalar field are given by,
\begin{equation}
\eta \sim Log(2 + \sqrt{3}) + a_\eta e^{(\triangle_{IR} - 4 ) r /L_{IR}}, \quad
A_t \sim a_{A_t} e^{(\triangle_{A_t} - 3) r/L_{IR}}.
\end{equation}
From the infra-red limit of the equations it follows that  $\triangle_{IR} = 6 - \sqrt{6}$ and $\triangle_{A_t} = 5 $.
The parameter $a_{A_t}$ can be chosen to be equal to 1 \cite{Gubser:2009gp} by shifting $r$. That would introduce a multiplicative factor in $e^{2A}$ in the metric, which can be reabsorbed by rescaling $t$ and $\vec{x}$ appropriately. So we are left with a single parameter $a_{\eta}$.

At ultraviolet, $\eta = 0$, $h = h_{UV}$, $A \sim \frac{r}{\sqrt{h_{UV}} L}$.
In order to ensure that the solution gives rise to spontaneous breaking of the symmetry it is required that at the ultraviolet  $\eta \sim e^{-3 A}$, which corresponds to an expectation value for the dimension 3 operator dual to $\eta$. Imposing this  condition allows only discrete values of $a_{\eta}$. With a suitable value of the parameter $a_{\eta}$, the equations (\ref{eom}) with boundary condition can be integrated numerically for domain wall solution.
 For the range that we have used we have chosen $a_\eta = 1.866$, which has least number of nodes. This solution is expected not to be supersymmetric and so there are possibilities of instabilities. An analysis of thermodynamic stability of the numerical solution is required to settle stability related issue. Our choice corresponds to the fact that, for other values of $a_\eta$ would give solutions with higher number of nodes with same boundary condition and so have higher free energy and can be considered as less favourable thermodynamically.   
 Profiles of the various fields are given in Fig.1. 

\begin{figure}[h]
			\centering
			\begin{subfigure}{7cm}
				\centering
				\includegraphics[width=7cm]{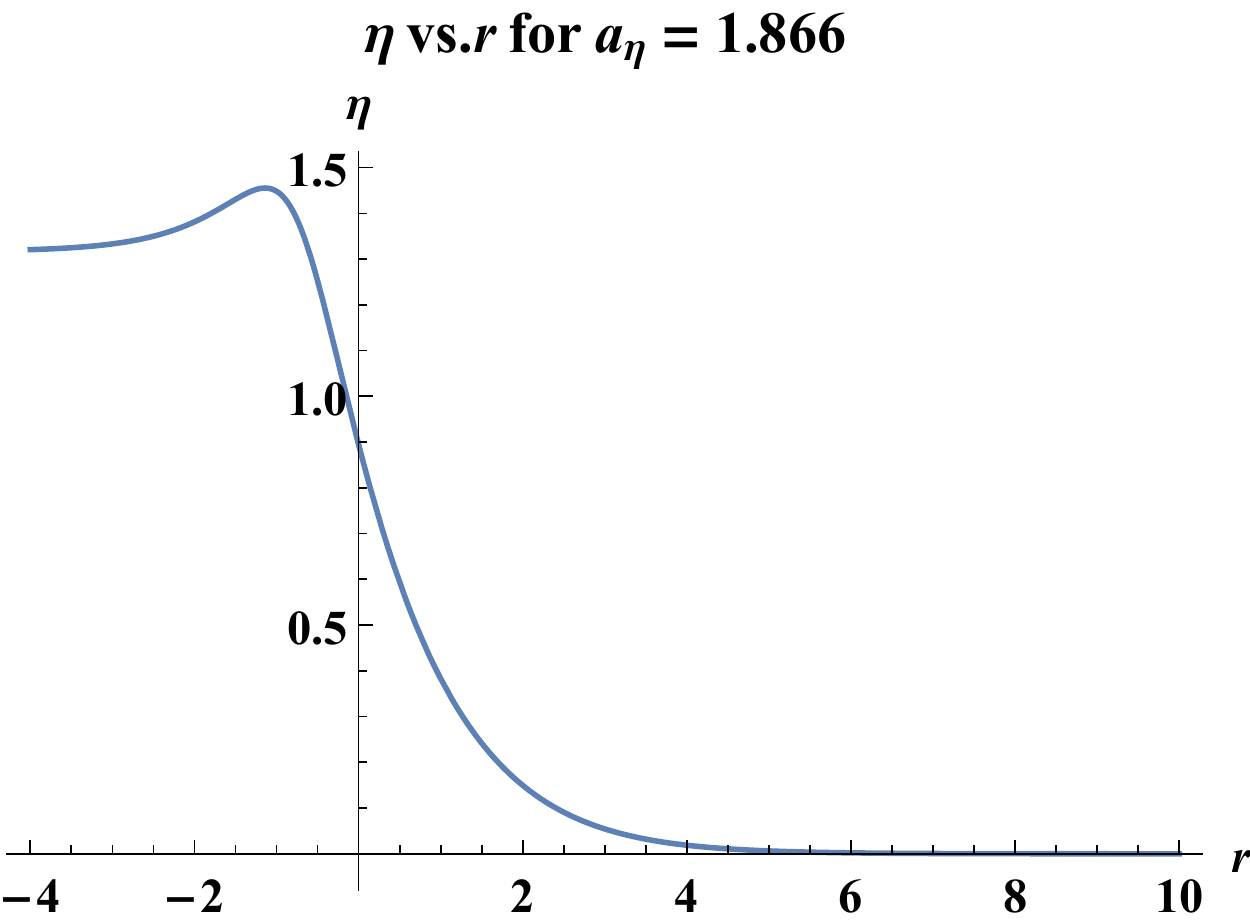}
				\caption{$\eta$ vs. $r$}
				\label{fig:sub1}
			\end{subfigure}%
			\begin{subfigure}{7cm}
				\centering
				\includegraphics[width=7cm]{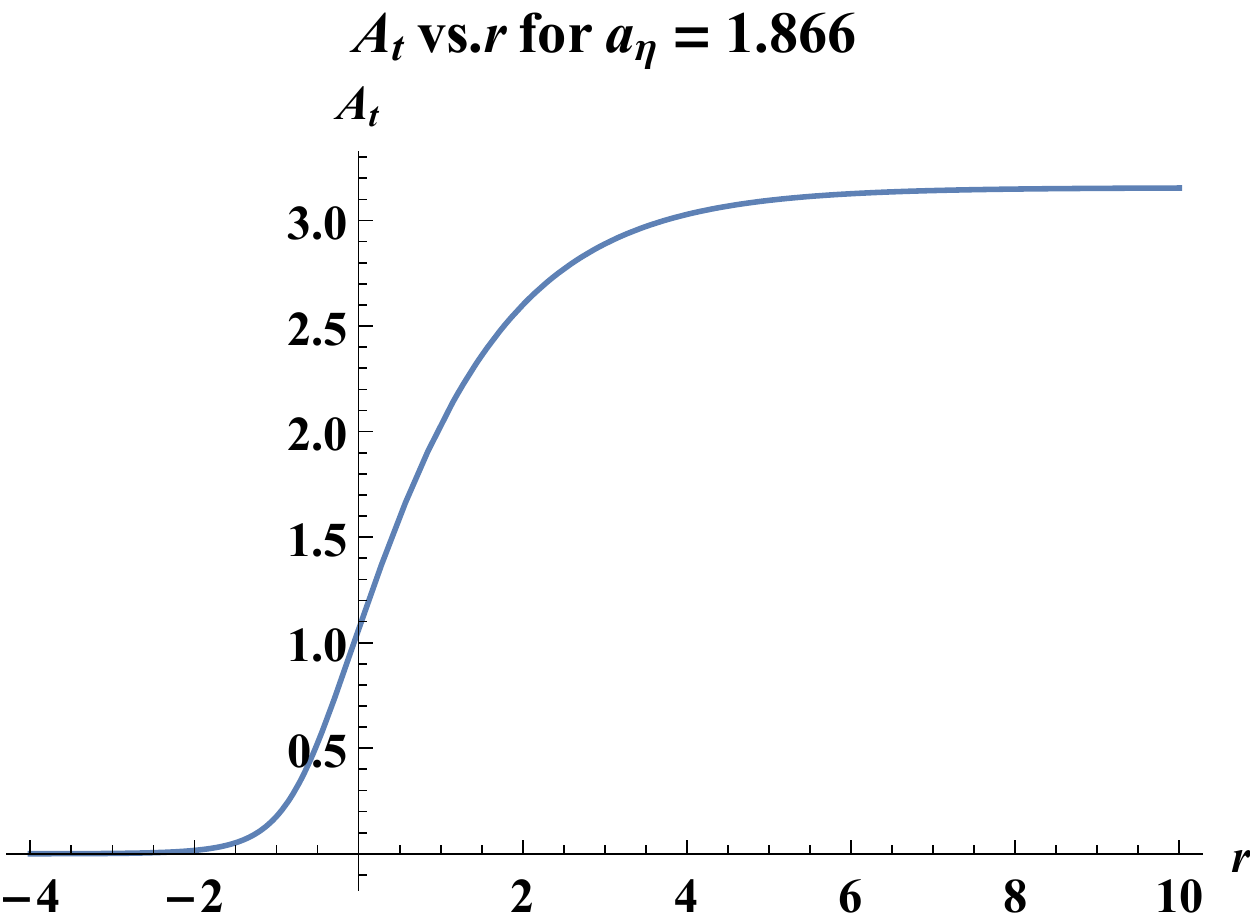}
				\caption{$A_t$ vs. $r$}
				\label{fig:sub2}
			\end{subfigure}
			\begin{subfigure}{7cm}
				\centering
				\includegraphics[width=7cm]{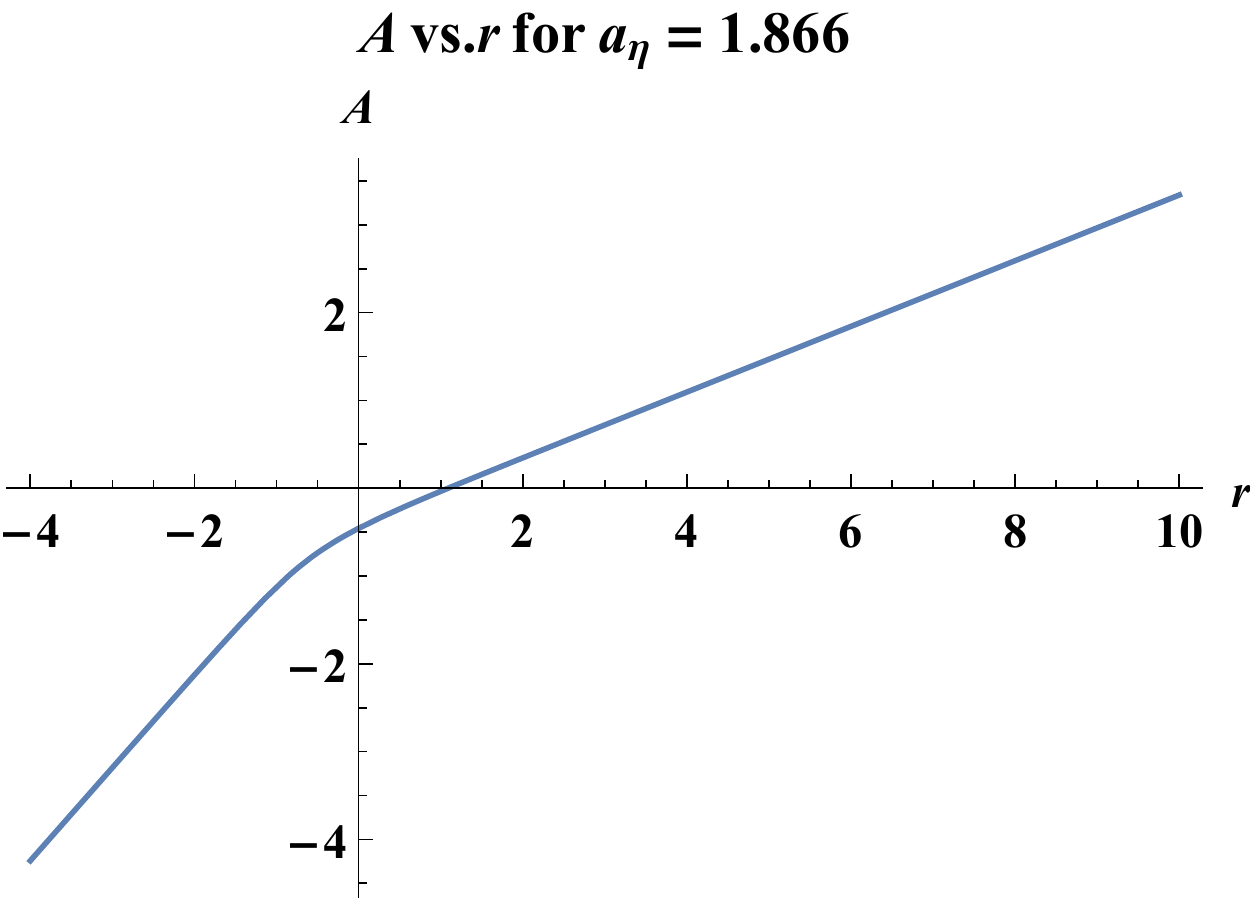}
				\caption{$A$ vs. $r$}
				\label{fig:sub1}
			\end{subfigure}%
			\begin{subfigure}{7cm}
				\centering
				\includegraphics[width=7cm]{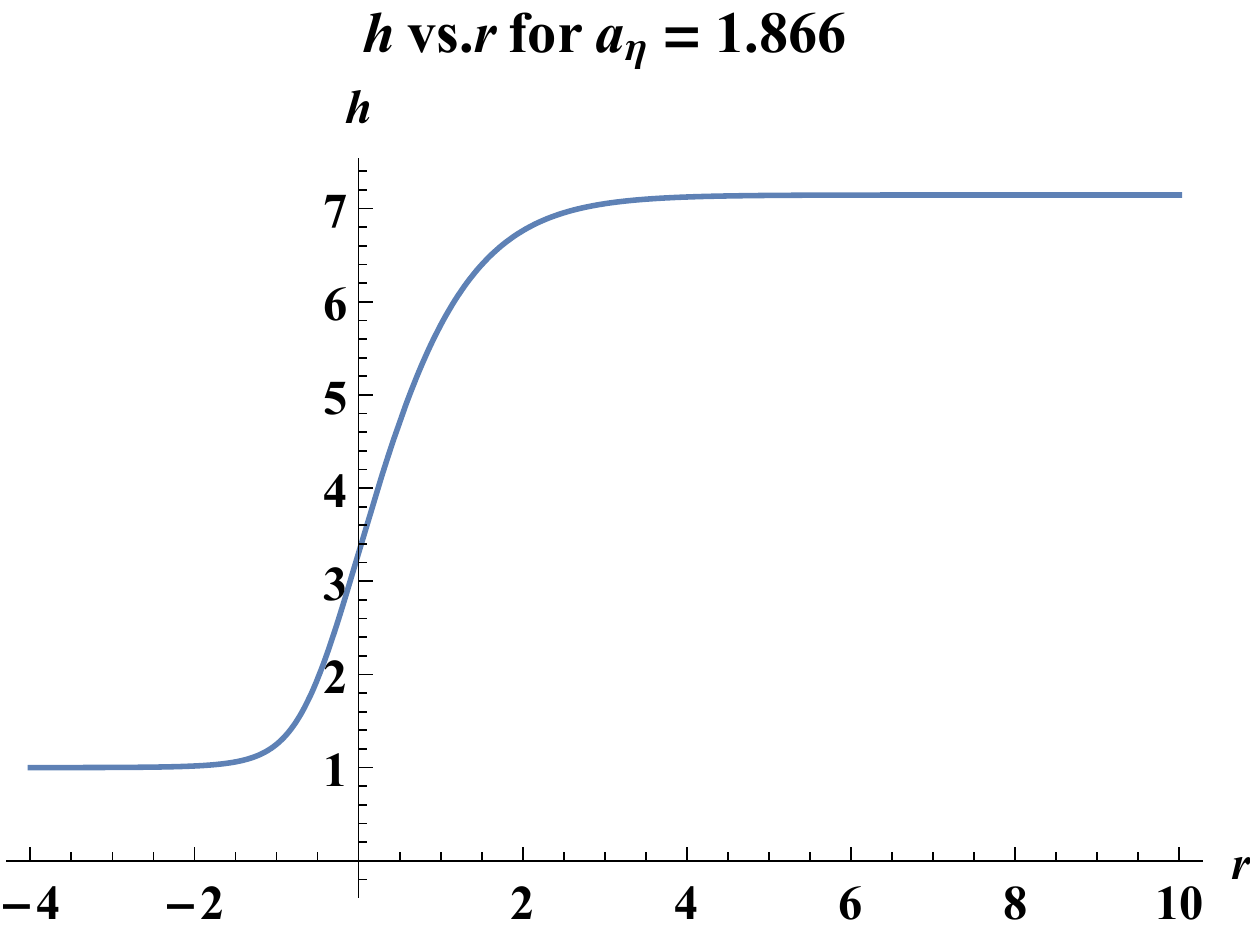}
				\caption{$h$ vs. $r$}
				\label{fig:sub2}
			\end{subfigure}
			\caption{Plots of different fields for domain wall solution}
			\label{fig:1}
		\end{figure} 

\section{Green's Function}

In this section we wil study the fermionic spectrum for the bosonic solution presented in the last section in the background. This domain wall appears as a solution in a consistent truncation of bosonic theory in a compactification of type IIB theory on AdS$_5$ times a Sasaki-Einstein manifold and the fermionic content of this truncated theory has been discussed elaborately in \cite{Bah:2010cu}. As explained there, after suitable truncation the fermionic fields can be arranged into separate decoupled sectors. For the present purpose we are interested in the sector consisting of only a single fermionic mode, which does not couple to any other fermionic mode or gravitino. That will keep the analysis simpler, while considering coupled fermions along with gravitino would require a more involved approach. The action for this sector containing a single fermionic mode is given by
\begin{equation}
S = \int ~d^5x ~ \sqrt{-g}~ \dfrac{1}{2} \bar{\lambda} (\Gamma^\mu D_\mu  + i\frac{\sqrt{3}}{2} \sinh^2\frac{\eta}{2} \Gamma^\mu A_\mu - \frac{1}{2}(7+\sinh^2\frac{\eta}{2}) - i p \frac{\sqrt{3}}{4} \Gamma^{\mu\nu}F_{\mu\nu} )\lambda ,
\end{equation}
where $ D_\mu = \partial_\mu + \frac{1}{4} \omega_{\mu ab}\Gamma^{ab} - i \frac{\sqrt{3}q}{2} A_\mu$ and $\omega_{\mu ab}$ represents the spin connection. We have set $L=1$. From the supergravity action the asymptotic charge $q=1$ and the coefficient of Pauli term $p=\frac{1}{6}$, but we have kept these as free parameter.
The Dirac equation following from the action is given by
\begin{equation}
[\Gamma^\mu (\partial_\mu - i \frac{\sqrt{3} Q}{2} A_\mu ) - M - i p \frac{ \sqrt{3}}{4} \Gamma^{\mu\nu} F_{\mu\nu} ] \lambda = 0,
\end{equation}
where $ Q = q + \sinh^2\frac{\eta}{2}$ and $M = \frac{1}{2}(7+\sinh^2\frac{\eta}{2})$ are the scaler dependent charge and mass terms  and so it has a running coupling and mass.  Redefining $\lambda \rightarrow h^{-1/8}$ as $\lambda$ we can absorb the contribution of spin connection in the Dirac equation.

 We choose the following $\gamma$-matrices in $2\times 2$ block form,
\begin{equation}
\Gamma^{\hat{t}} = \left(\begin{array}{cc}0& i\sigma_2\\i\sigma_2&0\end{array}\right),\quad
\Gamma^{\hat{r}} = \left(\begin{array}{cc}1& 0\\0&-1\end{array}\right),\quad
\Gamma^{\hat{x}} = \left(\begin{array}{cc}0& \sigma_1\\\sigma_1&0\end{array}\right).
\end{equation}
Spinors are chosen to be $\lambda = e^{- i \omega t + i k x }  (\psi^+ , \psi^- )^T$, where each of the $\psi^\pm$ are two component spinors. Dirac equations reduce to
\begin{equation}
(\pm \sqrt{h} \partial_r - M ) \psi^\pm + i e^{-A} ( k \sigma_1 - \frac{\omega + \frac{\sqrt{3} Q}{2}  A_t}{\sqrt{h}} i\sigma_2 \mp \frac{\sqrt{3} p}{2} A_t^\prime i\sigma_2 )\psi^\mp = 0.
\end{equation}
Writing $\psi^\pm = ( \psi^\pm_1 , \psi^\pm_2 )^T$, equations for the individual components becomes
\begin{equation}\begin{split}\label{diraceq1}
(\sqrt{h} \partial_r - M )\psi^+_1 + i e^{-A} [ k - \frac{\omega + \frac{\sqrt{3} Q}{2} A_t}{\sqrt{h}} - \frac{\sqrt{3} p}{2} A_t^\prime ] \psi^-_2 = 0,\\
(\sqrt{h} \partial_r + M )\psi^-_2 - i e^{-A} [ k + \frac{\omega + \frac{\sqrt{3} Q}{2} A_t}{\sqrt{h}} - \frac{\sqrt{3} p}{2} A_t^\prime ] \psi^+_1 = 0.
\end{split}\end{equation}
The other two components satisfy the following equations,
\begin{equation}\begin{split}
(\sqrt{h} \partial_r - M )\psi^+_2 + i e^{-A} [ k + \frac{\omega + \frac{\sqrt{3} Q}{2} A_t}{\sqrt{h}} + \frac{\sqrt{3} p}{2} A_t^\prime ] \psi^-_1 = 0,\\
(\sqrt{h} \partial_r +  M )\psi^-_1 - i e^{-A} [ k - \frac{\omega +  \frac{\sqrt{3} Q}{2} A_t}{\sqrt{h}} + \frac{\sqrt{3} p}{2} A_t^\prime ] \psi^+_2 = 0.
\end{split}\end{equation}
Note that ( $\psi^+_1$, $\psi^-_2$) and  ( $\psi^+_2$, $\psi^-_1$) are coupled with each other through the Dirac equation. Equations for these two sets will be interchanged  by flipping the signs of $\omega$, $Q$ and $p$. In what follows we will confine ourselves to the case of  ( $\psi^+_1$, $\psi^-_2$) only.

We consider the behaviours of the fermions following from (\ref{diraceq1}) at the IR and UV limits. At the IR limit, $\eta= Log(2 + \sqrt{3})$, which implies $Q= q + 1/2, m_{IR}= 15/4$, $h=1$ and the geometry is AdS with radius $L_{IR}$. Following \cite{Iqbal:2009fd}, the behaviour of fermions corresponding to in-falling boundary condition depends on whether the momentum is space-like or time-like. We discuss the two cases in the following separately.

We begin with space-like momenta, $k^2 \ge \omega^2$. For this case, in-falling boundary conditions at IR are given in terms of modified Bessel functions as follows:
\begin{equation}\begin{split}\label{bcspace}
\psi^+_1(r) \sim U^+_1 e^{- r / 2 L_{IR}} K_{m_{IR}L_{IR} + \frac{1}{2}} (\sqrt{k^2 - \omega^2} L_{IR} e^{-r/L_{IR}}),\\
\psi^-_2(r) \sim U^-_2 e^{- r / 2 L_{IR}} K_{m_{IR}L_{IR} - \frac{1}{2}} (\sqrt{k^2 - \omega^2} L_{IR} e^{-r/L_{IR}}),
\end{split}\end{equation}
where $U^-_2 = - i \sqrt{\frac{k+\omega}{k-\omega}}  U^+_1$. We have chosen $ U^+_1 = 1$.

For time-like momentum, $\omega| >  |k|$ the solutions are expressed in terms of Hankel function of first kind,
\begin{equation}\begin{split} \label{bctime}
\psi^+_1(r) \sim U^+_1 e^{- r / 2 L_{IR}} H^{(1)}_{m_{IR}L_{IR} + \frac{1}{2}} (\sqrt{\omega^2 - k^2} L_{IR} e^{-r/L_{IR}}),\\
\psi^-_2(r) \sim U^-_2 e^{- r / 2 L_{IR}} H^{(1)}_{m_{IR}L_{IR} - \frac{1}{2}} (\sqrt{\omega^2 - k^2} L_{IR} e^{-r/L_{IR}}),
\end{split}\end{equation}
where $U^-_2 =  i \sqrt{\frac{\omega + k}{\omega - k}}  U^+_1$. We have chosen $ U^+_1 = 1$. Similarly, for $\omega < - |k|$ they are expressed in terms of Hankel function of second kind.

For both the regions, at UV  limit, $\eta = 0$ and $Q=q$, $M_{UV}=7/2$, $h(r)$ approaches a constant $h_{UV}$, $A_t$ approaches $A_t(UV)$ and the geometry is AdS with radius $L_{UV}$. At $r\rightarrow \infty$ behaviour of fermions depend on mass terms only and are given by
\begin{equation}\begin{split}\label{asymptotic}
\psi^+_1(r) \sim C^+_1 e^{ M_{UV} r/\sqrt{h_{UV}}} +  D^+_1 e^{ - (M_{UV} + 1) r/\sqrt{h_{UV}}} ,\\
\psi^-_2(r) \sim C^-_2 e^{ (M_{UV} - 1) r/\sqrt{h_{UV}}} +  D^-_2 e^{ - M_{UV} r/\sqrt{h_{UV}}}.
\end{split}\end{equation}

The Green's function is given by
\begin{equation}\label{Green}
G_{R}(\omega, k) =
 \frac{  D^-_2}{ C^+_1}.
\end{equation}
The Green's function in the present case is diagonal and the other component can be obtained from  ( $\psi^+_2$, $\psi^-_1$) in a similar manner. Imaginary part of the retarded Green's function represents the spectral function. In the next section we study the behaviour of spectral function for fermions for different choices of charges.

\section{Result}

In this section we consider behaviour of the operators dual to the fermionic modes in the present model. As mentioned earlier, restricting ourselves to  ( $\psi^+_1$, $\psi^-_2$) is sufficient as the behaviour for the other two fermionic modes will be similar. Unlike generic fermions, in this model both charges and masses depend on the scalar field $\eta$ through the relation  $ Q = q - \sinh^2\frac{\eta}{2}$ and $M = \frac{1}{2}(7+\sinh^2\frac{\eta}{2})$.

Since the boundary conditions differ in spacelike and timelike region, these two cases are analysed separately. For the former (spacelike region) we numerically solve Dirac equations for supergravity fermionic modes (\ref{diraceq1}) subject to the boundary condition (\ref{bcspace}) and look for normal modes. The normal modes correspond to zeroes of $C_1^+$ in (\ref{asymptotic}) leading to singularities of the Green function. Keeping the charge $q$ fixed we scan over values of $\omega$ and $k$ to find the zeroes of $C_1^+$. We begin with the charge following from the supergravity model i.e. $q=1$, which does not yield any normal mode. This fermionic mode has small asymptotic charge ($q=1$) and non-zero asymptotic mass ($m=\frac{7}{2}$) and so it is consistent with the result in \cite{Gubser:2009dt}, as with higher mass possibility of having normal mode reduces. 

\begin{figure}[h]
			\centering
			\begin{subfigure}{7cm}
				\centering
				\includegraphics[width=7cm]{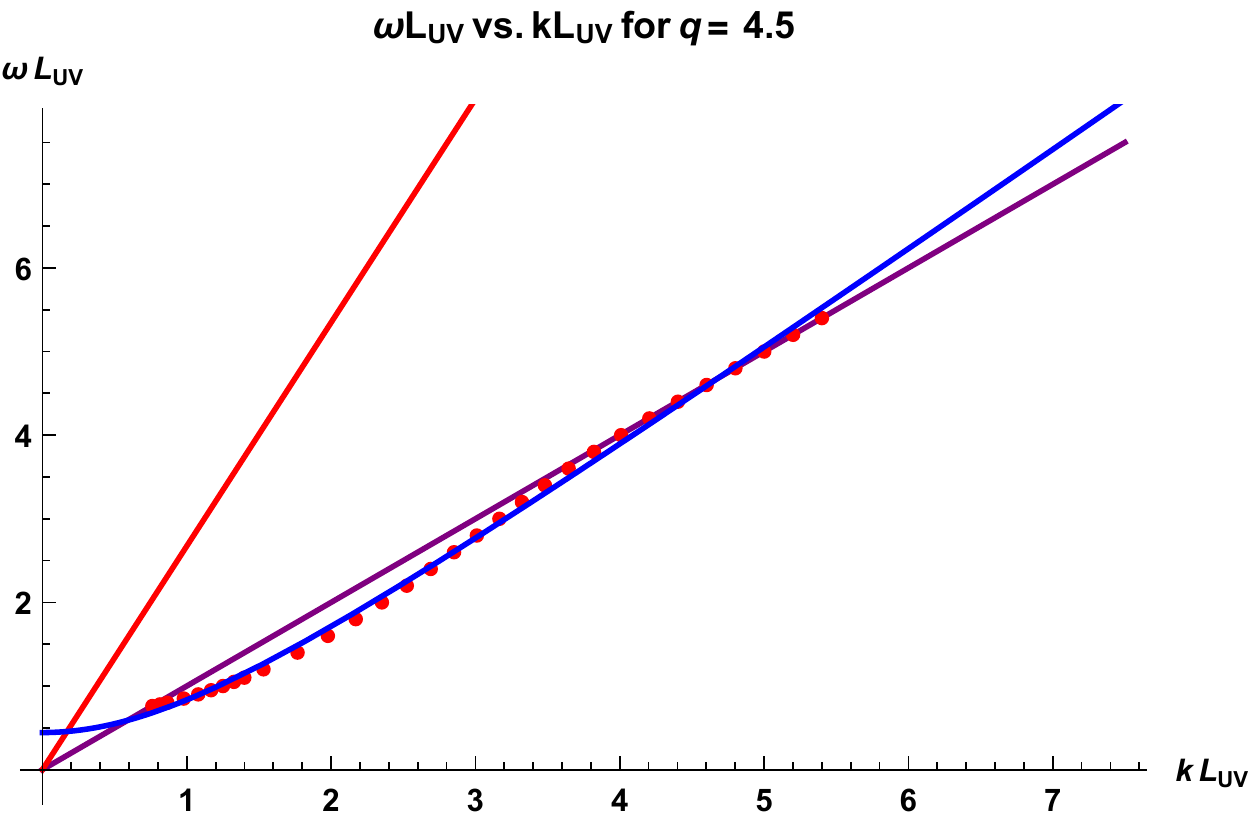}
				\caption{$\omega L_{UV}$ vs. $k L_{UV}$ for $q=4.5$}
				\label{bessel-normal45}
			\end{subfigure}%
			\begin{subfigure}{7cm}
				\centering
				\includegraphics[width=7cm]{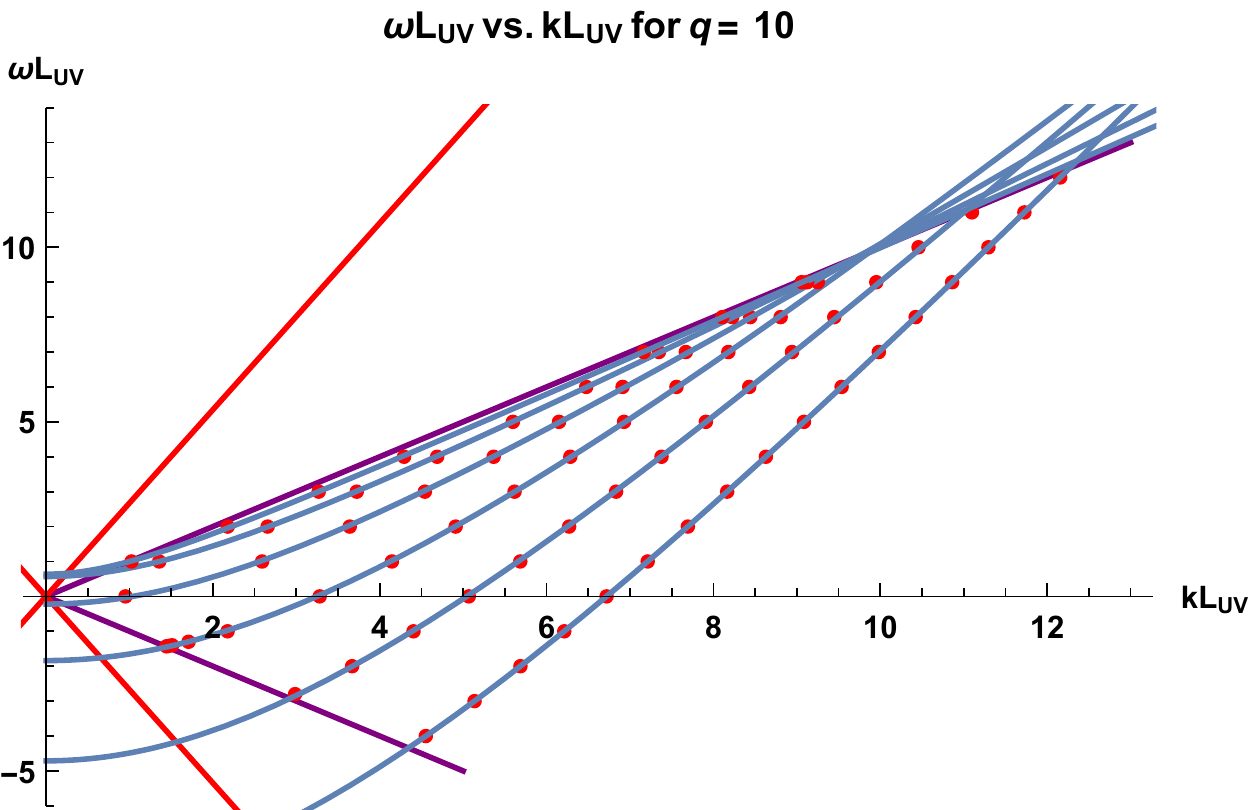}
				\caption{$\omega L_{UV}$ vs. $k L_{UV}$ for $q=10$ }
				\label{bessel-normal10}
			\end{subfigure}
			\caption{Normal modes in the space-like region for $q=4.5$ and $q=10$. The solid purple lines   and red lines represents boundaries of IR and UV lightcones respectively. Blue lines show the fits. }
			\label{normal}
		\end{figure}

Artificially dialling the charge to higher values leads to the appearances of normal modes for $\omega \geq \omega_c $ in this region. As charge increases $\omega_c$ decreses coming down to $\omega_c=0$.
This has been demonstrated  for two different values of charges, $q=4.5$ and $q=10$.  Unless mentioned otherwise, we have kept the mass and Pauli coupling to be same as those followed from supergravity throughout the discussion. The plots of $\omega$ vs. $k$ for normal modes for those two charges are given in Fig.\ref{bessel-normal45} and  Fig.\ref{bessel-normal10} respectively. For $q=4.5$ normal modes occur for $\omega \geq \omega_c=0.761$, while for $q=10$ the minimum value for $\omega$ for occurance of normal modes comes down to $\omega_c=0$ indicating gapped and gapless spectra in these two cases respectively. 

\begin{figure}[h]
			\centering
			\begin{subfigure}{7cm}
				\centering
				\includegraphics[width=7cm]{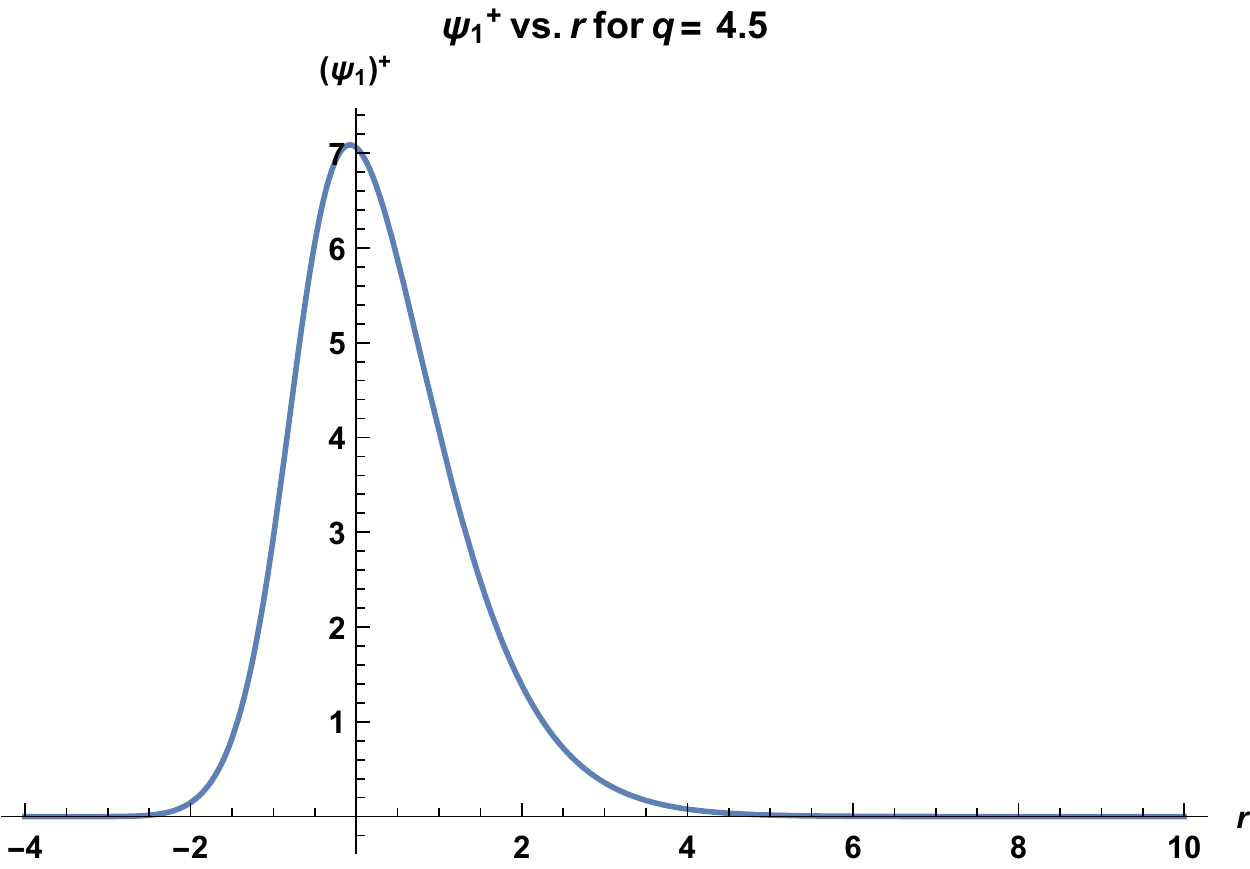}
				\caption{$\psi^+_1$ vs. $r$ for $q=4.5$ for a typical normal mode.}
				\label{bessel-normal-wf}
			\end{subfigure}%
			\begin{subfigure}{7cm}
				\centering
				\includegraphics[width=7cm]{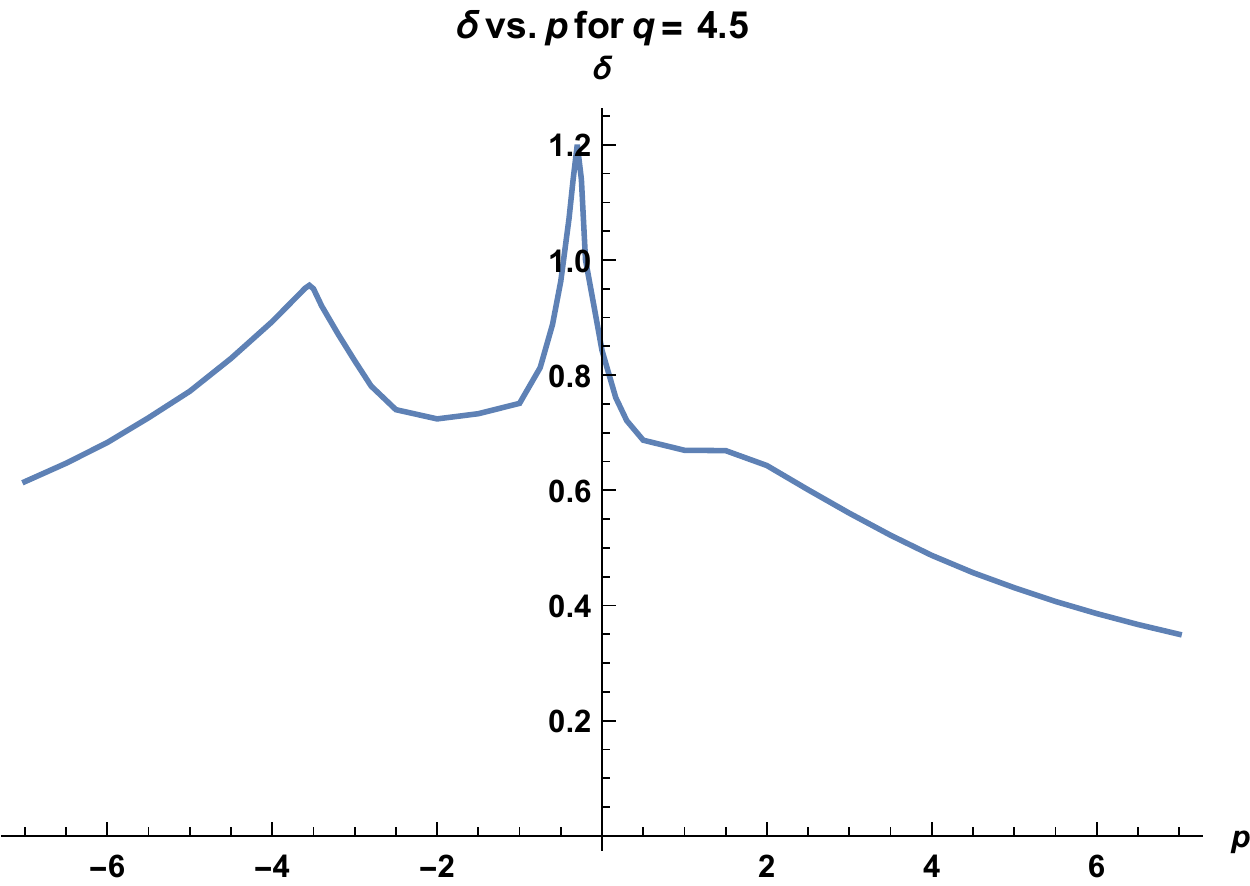}
				\caption{Gap vs. Pauli term coefficient in spacelike region for $q=4.5$}
				\label{gap}
			\end{subfigure}
			\caption{ }
			\label{wvfn}
		\end{figure}

In the present model, the supergravity Lagrangian has a Pauli term with coefficient $p=\frac{1}{6}$ and as shown in \cite{Edalati:2010ww,Wu:2014rqa} Pauli term may have substantial effect on the spectrum. In particular, as observed in  \cite{Wu:2014rqa}, large Pauli term may give rise to gapped spectrum. In order to explore such effects in our model we have manually varied the coefficient of the Pauli term $p$ keeping the charge $q$  fixed at $4.5$ and plotted the gap $\delta$ vs. $p$ in Fig \ref{gap}.
We find for a small negative value of $p$ (around -0.3) the gap is maximum. As we go away from this value on both sides the gap generally decreases, apart from a local maxima around $p=- 3.55$. Since in the present case, the gap is non-zero at $p=0$ for smaller charge, it cannot be interpreted as an effect of Pauli term.

It has been suggested from a semi-classical analysis\cite{Gubser:2009dt}, that the dispersion relation satisfied by the normal modes can be given by $\frac{(\omega + q \phi_{UV} )^2}{h_{UV}} - k^2 = m_{eff}^2$, where the constant on the right hand side is related to the number of nodes of the fermion wave-function associated with normal mode. For $q=4.5$ the normal modes shown in Fig.\ref{bessel-normal45} correspond to wavefunctions with zero nodes. We have plotted a typical wavefunction in Fig. \ref{bessel-normal-wf}. As we increase the value of charge $q$ to $q=10$, more normal modes appear. These are organised along various curves shown in  Fig.\ref{bessel-normal10}, where associated wavefunctions of the modes lying on a curve have same number of nodes. In Fig.\ref{bessel-normal10}, the normal modes lying on the curve on right extreme correspond to zero node wavefunctions and number of nodes increases as one moves from right to left. We have tried fitting the relation $\frac{(\omega + q \phi_{UV} )^2}{h_{UV}} - k^2 = m_{eff}^2$ with the points, but a one parameter fit, keeping the values of $ \phi_{UV}$ and $h_{UV}$ as follows from the equations and varying $m_{eff}$ does not reproduce the shapes of the curves well. Instead we have tried a 3 parameter fit by varying  $ \phi_{UV}, h_{UV}$ and $m_{eff}$ as arbitrary parameters and the shapes are reproduced well, as given in the  Fig.\ref{bessel-normal45} and  Fig.\ref{bessel-normal10}.
Introducing  $k_{UV}^2 = k^2 -  \frac{(\omega + q A_t(UV))^2}{h_{UV}}$ we find that, all the normal modes appear inside the region $ k \leq  \frac{(\omega + q A_t(UV))^2}{h_{UV}}$ as found in \cite{Gubser:2009dt}.

 
\begin{figure}[h]
			\centering
			\begin{subfigure}{7cm}
				\centering
				\includegraphics[width=7cm]{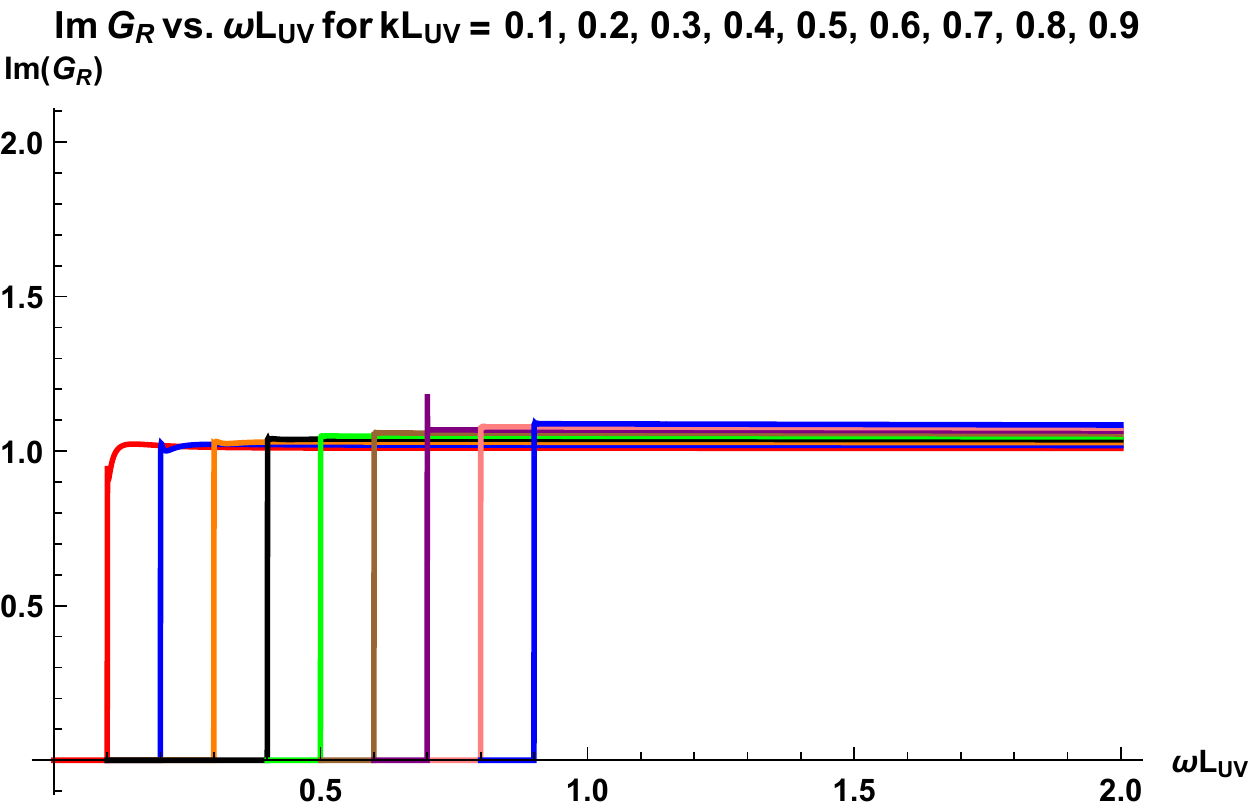}
			\caption{Spectral function for $q=10$, $\omega > k$}
			\label{masslessL}
			\end{subfigure}%
			\begin{subfigure}{7cm}
				\centering
				\includegraphics[width=7cm]{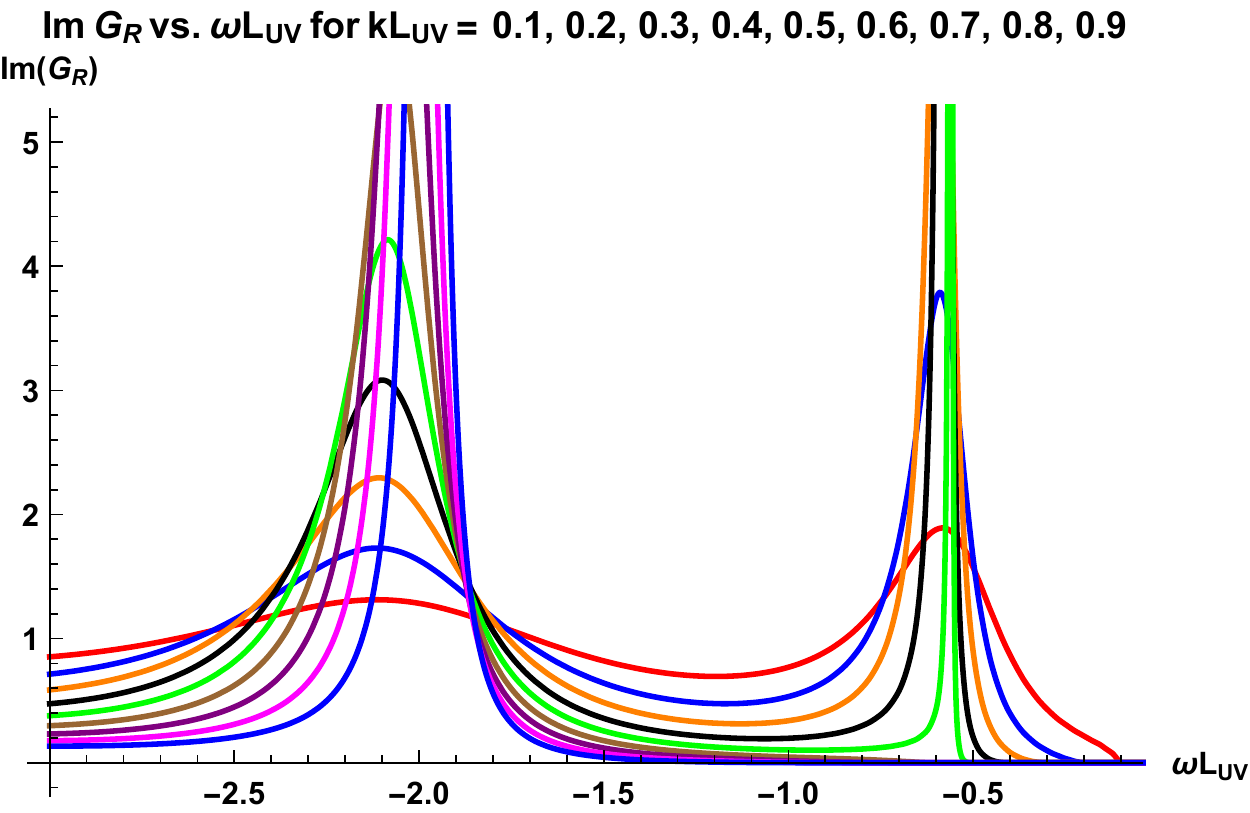}
				\caption{Spectral function for $q=10$, $\omega < -k$}
				\label{masslessR}
			\end{subfigure}
			\caption{Spectral function for fermionic mode in absence of mass term. $kL_{UV}$ = 0.1(red), 0.2(blue), 0.3(orange), 0.4(black), 0.5(green), 0.6(brown), 0.7(purple), 0.8(pink), 0.9(blue) }
			\label{massless}
		\end{figure}

Next we consider the complementary timelike region and explore the behaviour of spectral function by numerically solving Dirac equation (\ref{diraceq1}) with boundary condition given by (\ref{bctime}). The spectral function is obtained from imaginary part of the Green's function given in (\ref{Green}). We begin with Dirac equation in absence of the mass term and Pauli term for charge $q=10$ and plot the spectral function vs. $\omega$ for different values of $k$ are shown  in Fig.s \ref{masslessL} and  \ref{masslessR}.
%
 In order to find dispersion relation for these excitations we have plotted $\omega$ and $k$ values for the peaks  for  $k<0$ in Fig.\ref{dispersion0}. As it is evident from the figure the set of points in the uv region smoothly connect to the points representing normal modes residing in the IR region on the other side of the IR light cone (purple line). A numerical fit with expression  $\frac{(\omega + q \phi_{UV} )^2}{h_{UV}} - k^2 = m_{eff}^2$  by varying all the three parameters captures the shape of the curve well. One can obtain a similar plot for $k>0$. 

\begin{figure}[h]
			\centering
			\begin{subfigure}{7cm}
				\centering
				\includegraphics[width=7cm]{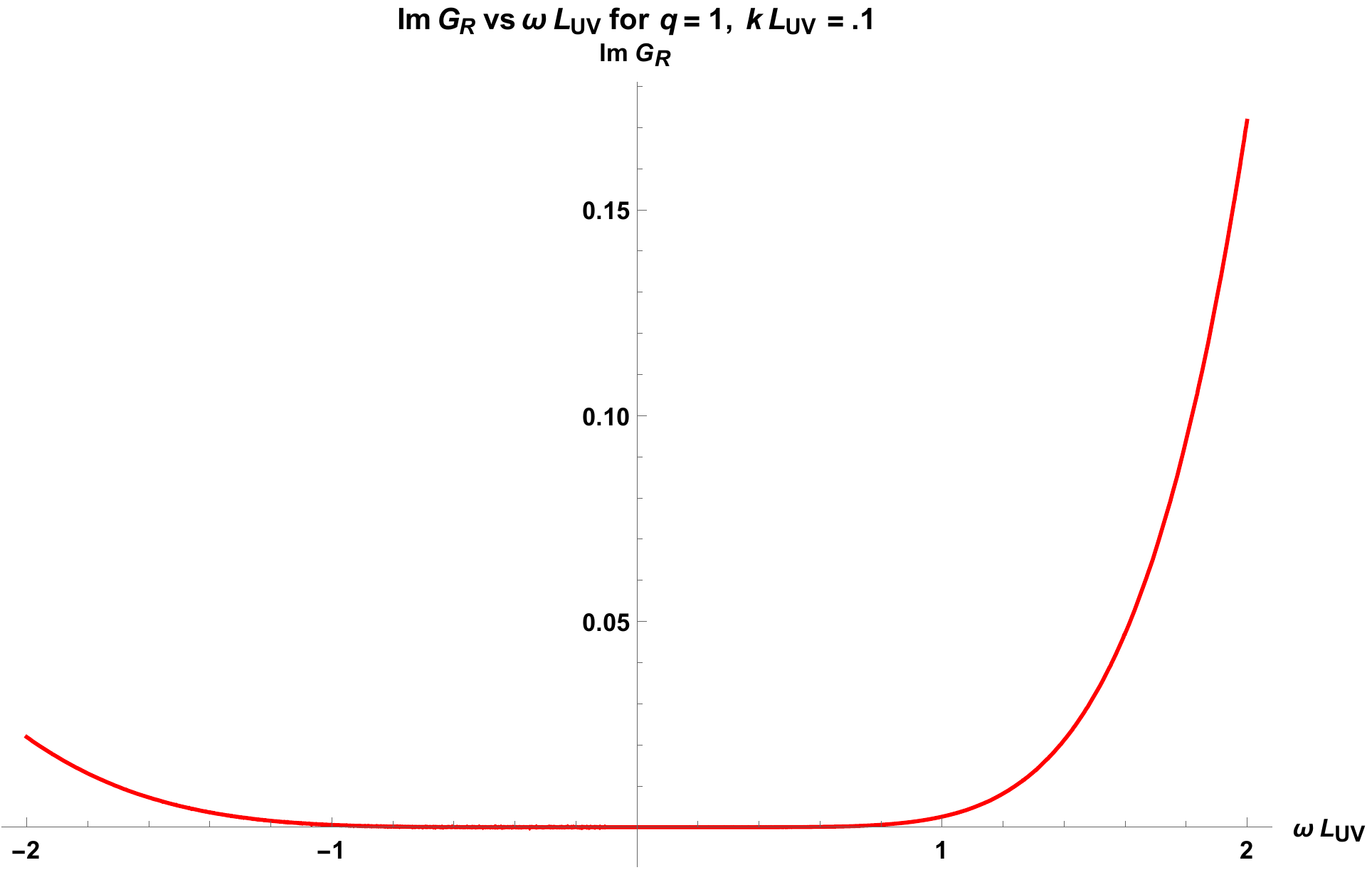}
			\caption{Spectral function for $q=1$; $kL_{UV}=0.1$ }
			\label{q1}
			\end{subfigure}%
			\begin{subfigure}{7cm}
				\centering
				\includegraphics[width=7cm]{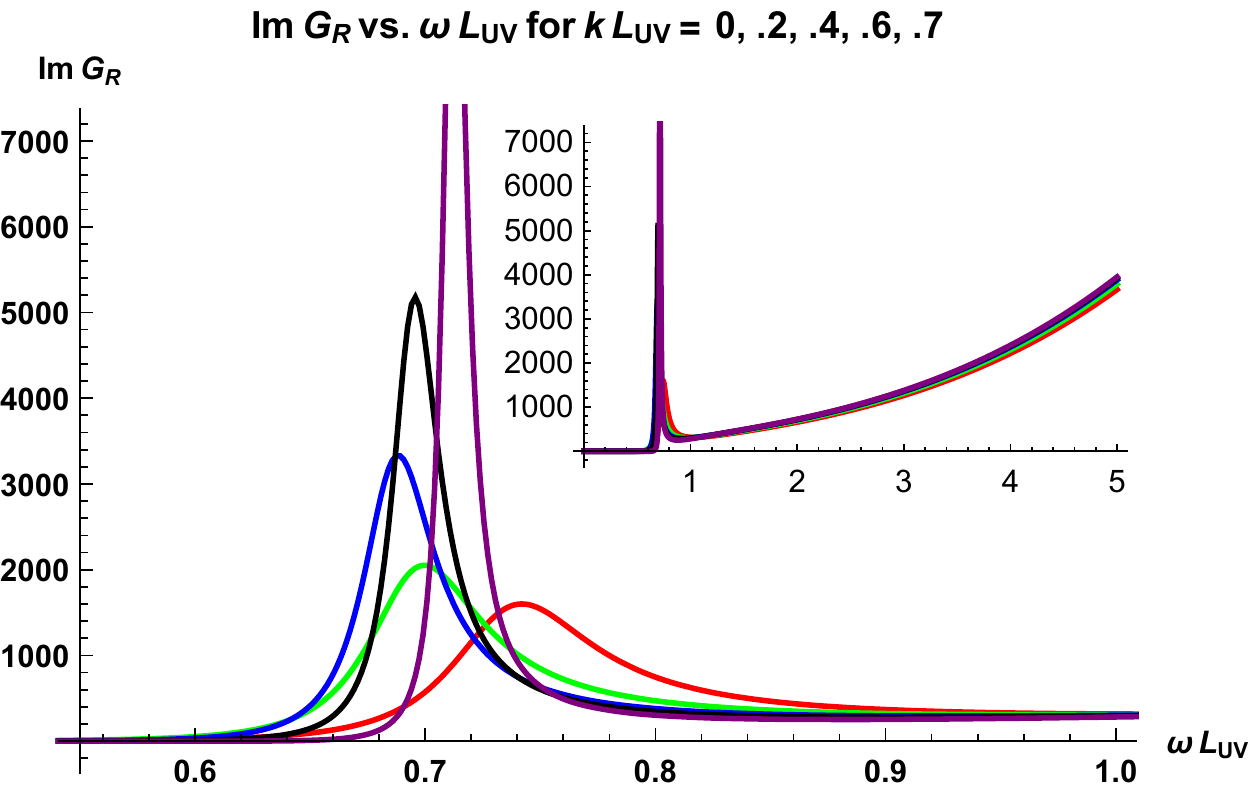}
				\caption{$Im G_R$ vs. $\omega L_{UV}$ for $q=4.5$}
				\label{hankel45}
			\end{subfigure}
			\caption{Spectral function for fermionic mode. Left $q=1$. Right $q=4.5$ with $kL_{UV}$=0 (red), 0.2(green), 0.4(blue), 0.6(purple) and 0.7(brown). The inset figure shows the behaviour at large frequency. }
			\label{hankel1}
		\end{figure}

\begin{figure}[h]
			\centering
			\begin{subfigure}{7cm}
				\centering
				\includegraphics[width=7cm]{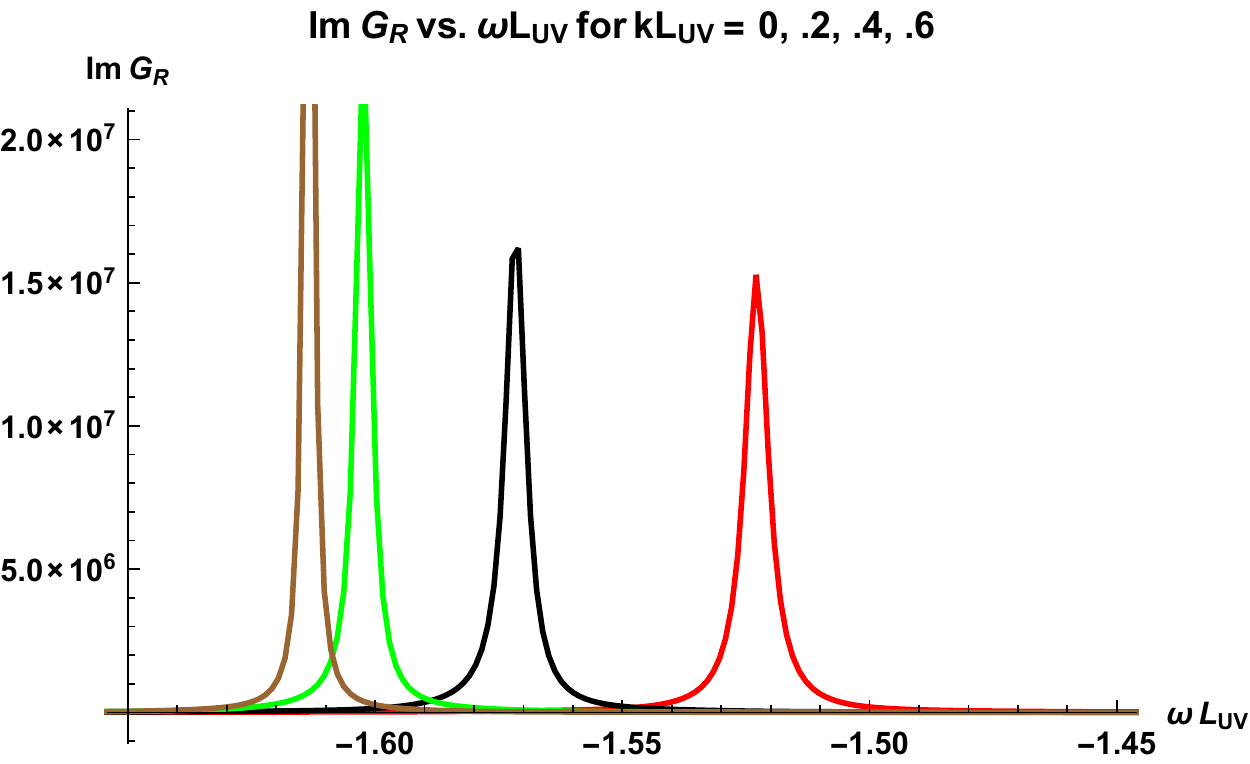}
				\caption{$Im G_R$ vs. $\omega L_{UV}$ for $q=10$  }
				\label{hankel10}
			\end{subfigure}%
			\begin{subfigure}{7cm}
				\centering
				\includegraphics[width=7cm]{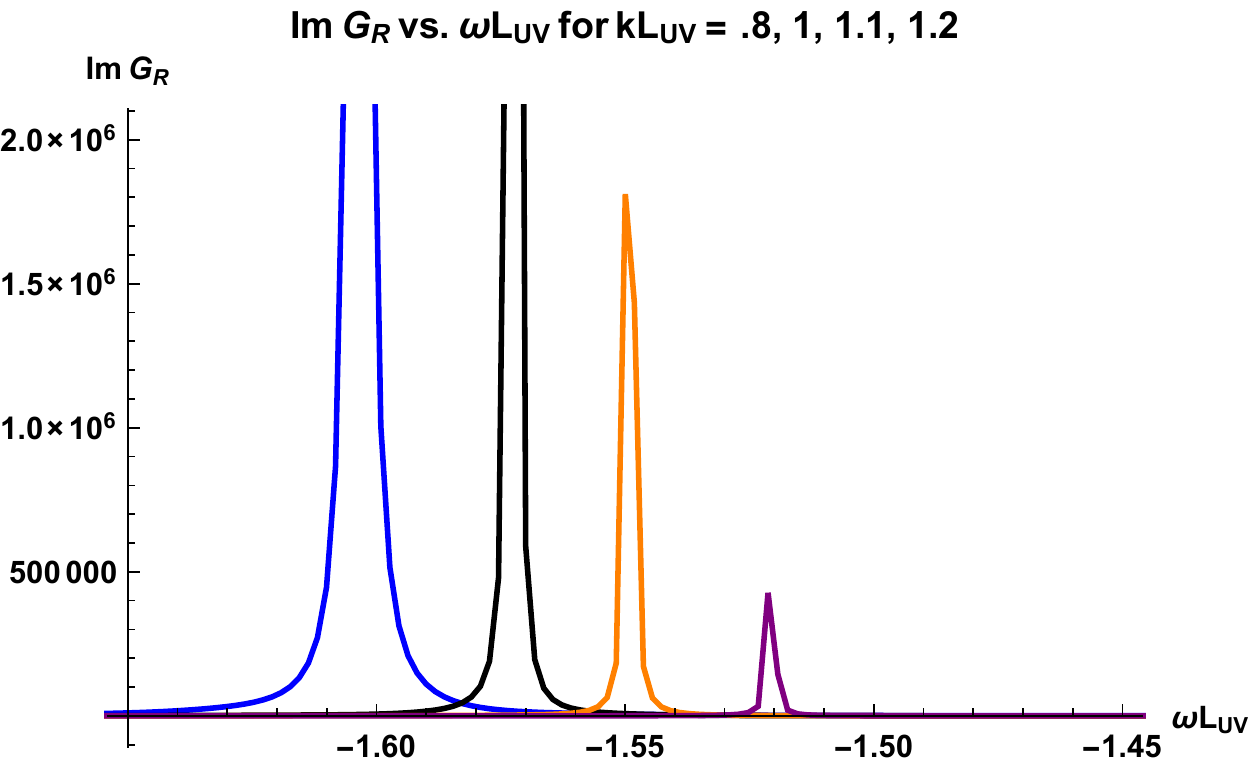}
				\caption{$Im G_R$ vs. $\omega L_{UV}$ for $q=10$}
				\label{hankel2}
			\end{subfigure}
			\caption{Spectral function for fermionic mode for $q=10$. On left with $k L_{UV}$=0(red), .2(black), .4(green), .6(brown). On right with $k L_{UV}$=.8(blue), 1(black), 1.1(orange), 1.2(purple). }
			\label{hankel}
		\end{figure}

\begin{figure}[h]
		 \centering
			\begin{subfigure}{7cm}
				\centering
				\includegraphics[width=7cm]{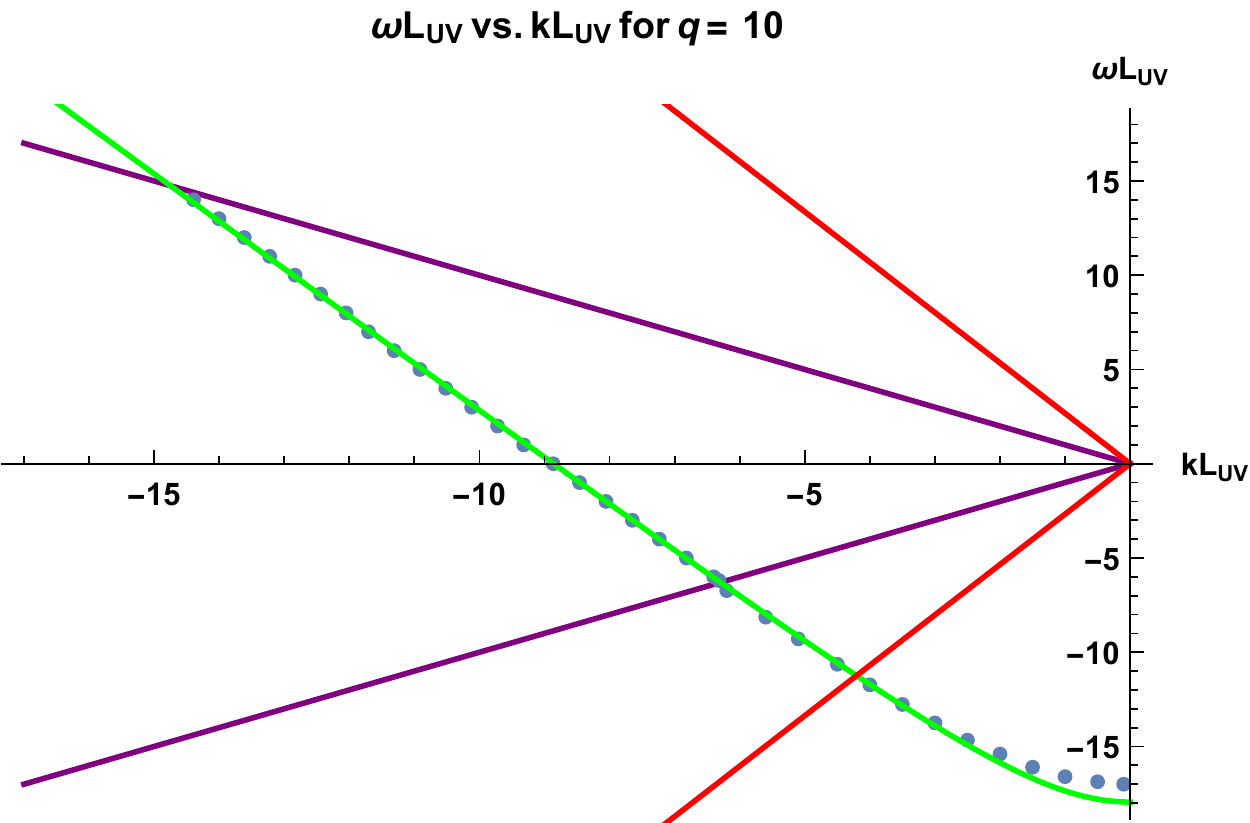}
				\caption{Without mass and Pauli term  }
				\label{dispersion0}
			\end{subfigure}%
			\centering
			\begin{subfigure}{7cm}
				\centering
				\includegraphics[width=7cm]{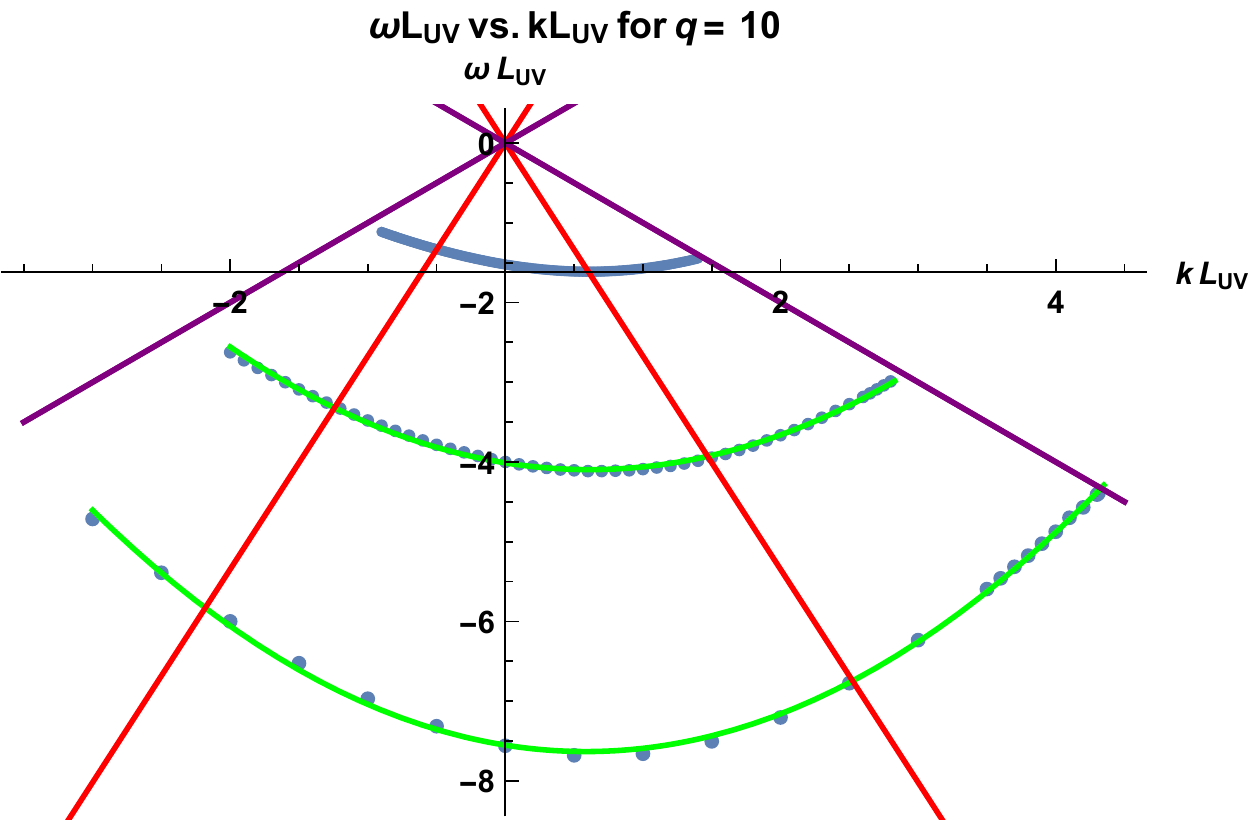}
			\caption{With mass and Pauli term. }
			\label{dispersion2}
			\end{subfigure}%
			\caption{Dispersion relation for $q=10$ in time-like region. The solid purple lines and red lines represent boundaries of IR and UV lightcones respectively. Green lines show the fits. Left figure shows  all regions in $k<0$. Right figure shows timelike region. }
		\end{figure}

\begin{figure}[h]
			\centering
			\begin{subfigure}{7cm}
				\centering
				\includegraphics[width=7cm]{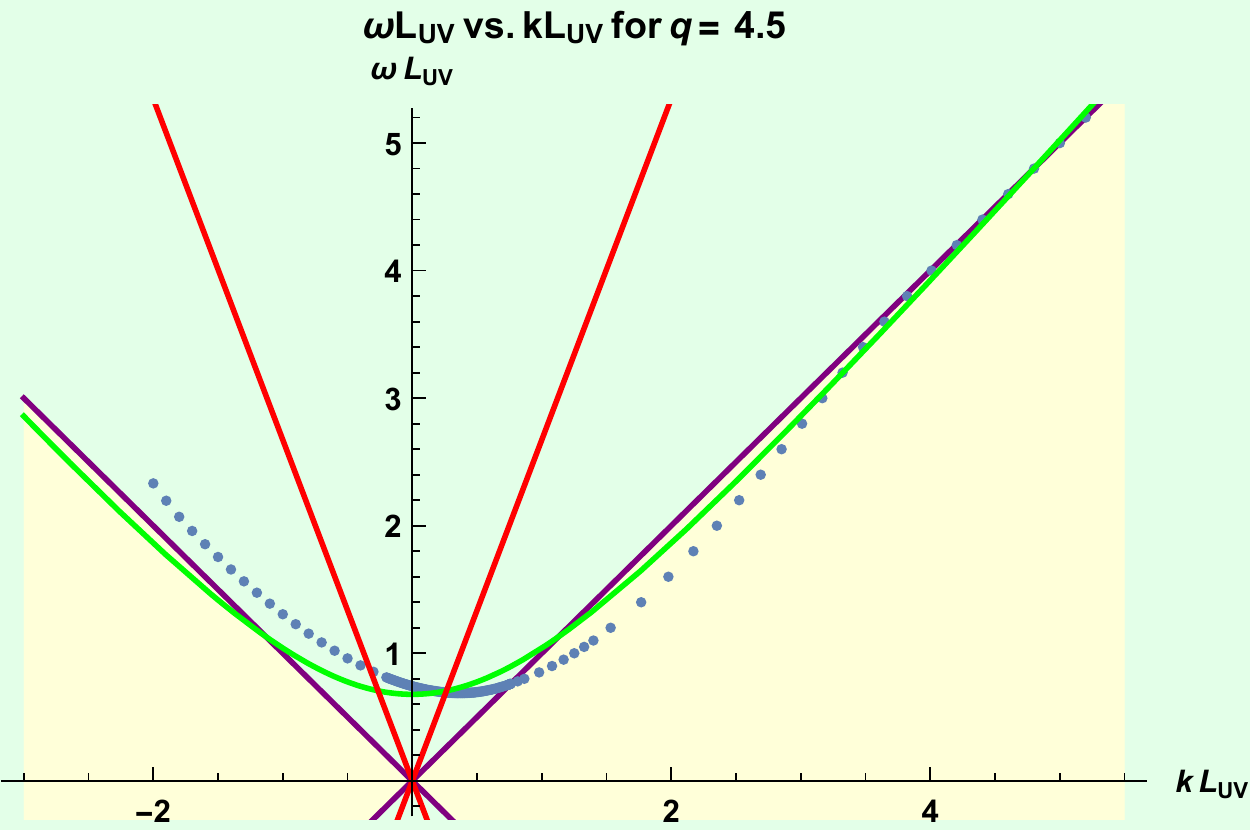}
			\caption{Dispersion relation for $q=4.5$ }
			\label{disp45}
			\end{subfigure}%
			\begin{subfigure}{7cm}
				\centering
				\includegraphics[width=7cm]{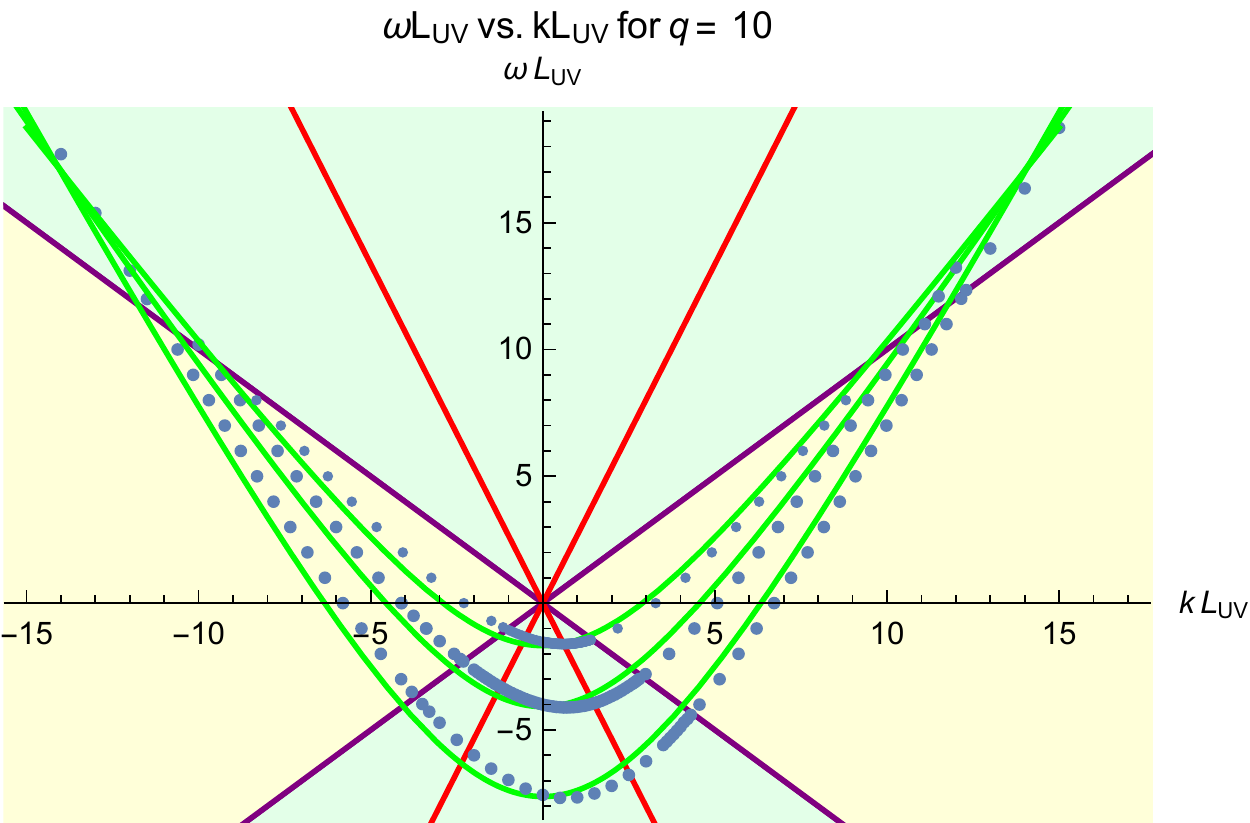}
				\caption{Dispersion relation for $q=10$}
				\label{disp10}
			\end{subfigure}
			\caption{Spectral function for fermionic mode for $q=4.5$ and $q = 10$. The solid purple lines and red lines represent boundaries of IR and UV lightcones respectively. Green lines show the fits. }
			\label{dispersionAll}
		\end{figure}

Next we consider the cases with both mass term as well as Pauli terms in place for two different values of charges, $q=4.5$ and $q=10$ and plot the spectral function vs. $\omega$ for  five and eight different values of $k$ respectively. The plots are given in Fig.\ref{hankel45} for $q=4.5$ and in Fig.\ref{hankel} for $q=10$. As the charge increases the heights of the peaks also increases. However, as $k$ increases the position of peak in $\omega$ does not vary monotonically. 
The dispersion relation of the excitations associated with these peaks can be observed from the plot of the positon of the peaks in $\omega$ vs. $k$. For modes within the time-like region for $q=10$ are shown in Fig.\ref{dispersion2}. Fitting the numerical data with  $\frac{(\omega + q \phi_{UV} )^2}{h_{UV}} - k^2 = m_{eff}^2$ by varying all the 3 parameters $\phi_{UV}$,  $h_{UV}$ and $m_{eff}$, does not yield a suitable fit. Instead, a quadratic fit,  with a relation like $\omega - \omega_0 =  \frac{(k - k_0)^2}{2 m_{\text{eff}}}$ matches with the data points in this region, as shown in the figure \ref{dispersion2}. Similar matches are obtained for $q=4.5$ 
 Fig.\ref{dispersionAll} shows that for both $q=4.5$ and $q=10$ the modes appearing in the timelike region matches smoothly with those corresponding to the normal modes in the space-like region. For $q=4.5$ the modes are trailing along the boundary of IR lightcone for positive $k$, while for $q=10$ modes for large frequency appears in the timelike region. However, in both the cases, for timelike region the modes are outside the UV lightcone. Considering both the regions, a 3 parameter numerical fit with relation given by $\frac{(\omega + q \phi_{UV} )^2}{h_{UV}} - k^2 = m_{eff}^2$ (shown by the green lines in the figure) is more suitable. It would be interesting to understand the features of such excitations in a greater detail. 
 In addition, from this Fig.\ref{dispersionAll} one can observe that for $q=4.5$ the modes are lying on the positive $\omega$ region with an upward concave pattern, which is expected for a BCS superconductor. For $q=10$, however, a similar pattern is obtained in the negative $\omega$ region, which is different from the one obatined for gapped spectrum in \cite{DeWolfe:2016rxk}. It could be due to the fact that for large values of charge the modes are shifted downwards in $\omega$.
For the charge $q=1$, which follows from the supergravity, however we have not found any peak as shown in Fig.\ref{q1}. This may be attributed to the fact that charge of this mode is too small. However, the large frequency behaviour are similar for other charges (for $q=4.5$ an inset figure in Fig.\ref{hankel45} is given to show this bahaviour).

		
We conclude this section with a discussion of the dual field theory. The dual model of this five dimensional supergravity theory, that we have considered corresponds to a four dimensional superconformal quiver gauge theory. The scalar field $\eta$ has mass given by $m^2=\triangle(\triangle -4)=-3$ implying conformal dimension of the dual operator is $\triangle =3$ with R-charge 2, confirming it is chiral primary. For IIB theory compactified on $S^5$ there are two such operators given by superpotential ${\mathcal W}$ and $tr(W_\alpha W^\alpha)$, where $W_\alpha$ is field strength superfield. As explained in \cite{Ceresole:1999zs}, only a linear combination of these two operators (orthogonal to chiral superfield associated with Konishi multiplet) represents the chiral primary. Following \cite{Gubser:2009qm} we identify the dual operator ${\mathcal O}_\eta$ as the lowest component of a linear combination of these two.

However, it turns out \cite{Gubser:2009qm} that for a black hole background, temperature of condensation of chiral primaries is a monotonically decreasing function of conformal dimension $\triangle$. Since the present case corresponds to zero temperature it may be useful to check that whether ${\mathcal O}_\eta$ has lowest conformal dimension compared to other chiral primaries in the dual theory. For IIB on $S^5$, the first family of scalar fields \cite{Oz:1998hr} admits modes with  $m^2=k(k-4), k\geq 2$, which couple to $\triangle = k$ chiral primary operators given by symmetrised traceless combinations $tr(\Phi^{i_1}...\Phi^{i_k})$. For $k=2$ this operator has conformal dimension $\triangle=2$ which is less than that of ${\mathcal O}_\eta$. Similarly, chiral primary operators with $\triangle <3$ exists in IIB on $T^{1,1}$  \cite{Ceresole:1999zs}, where $tr(A_iB_j)$ is a chiral primary with $\triangle = 3/2 < 3$. A suitable option would be to consider IIB on an orbifold of 5-sphere, $S^5/\Gamma$, where $\Gamma \in SU(3)$ and the dual theory is ${\mathcal N}=1$ supersymmetric quiver gauge theory. In the case of $\Gamma=Z_3$ orbifold, chiral primaries are discussed in  \cite{Oz:1998hr}. As explained there, the supergravity mode corresponding to $k=2$ mentioned above, is in $20'$ of $SU(4)$. For $\Gamma = Z_3$, its decomposition under $SU(3)\times U(1)$ is $20' = 6(4/3) + \bar{6}(-4/3) + 8(0)$. Only the $8(0)$ survives the projection but it does not couple to a chiral primary operator and so one does not expect to have a chiral primary of dimension 2 in the dual theory. However, superpotential and $tr(W_\alpha W^\alpha)$ will survive the orbifolding making ${\mathcal O}_\eta$, chiral primary operator with lowest dimension.  		
		
The fermions considered in \cite{Bah:2010cu} corresponds to the lowest rung of mass spectrum of fermions obtained by compactification of IIB on $S^5$ as given in Fig.4 of \cite{Kim:1985ez}. In particular, the fermionic mode we are interested in corresponds to mass $\frac{7}{2}$, that occurs in ${\mathbf 4}$ of $SU(4)$. In the notation of \cite{Gunaydin:1984fk,Ardehali:2013gra} it occurs at the level 3 sets of modes in representation $D(p+5/2, 1/2,0;0,p-3,1) +  D(p+5/2, 0,1/2;1,p-3,0)$  at $p=3$. The surviving KK modes for IIB on $S^5/{\mathbb Z}_3$ has been discussed and classified in  \cite{Ardehali:2013gra}.  Under $SU(3)\times U(1)$ decomposition the Dynkin label splits into $(1,p-3,0) = \oplus_{l=(p+1)/3}^{(2p-1)/3} (-1 + 2l -p, -1 + 2p -3l)_{2p-4l+1}$ $\oplus_{l=(p/3+1}^{(2p/3} (-3 + 3l -p, 0 + 2p -3l)_{2p-4l+1} $. The mode we are interested in corresponds to $p=3$ and in the second term in the splitting in this series and gives rise to a singlet of $SU(3)$. As explained in \cite{Ardehali:2013gra} it belongs to Gravitino multiplet II  ($\lambda^4$ in their notation). The corresponding superfield in the dual theory is given by $L_{2\dot{\alpha}} = tr (e^V \bar{W}_{\dot{\alpha}} e^{-V} W^2)$, where $V$ is the gauge superfield for the dual quiver gauge theory and $W$ represents the field strength superfield.  

\section{Discussion}

We have considered a domain wall solution with asymptotic AdS geometry that appears in a five dimensional supergravity theory obtained through compactification on a Sasaki-Einstein manifold. The dual theory is a quiver gauge theory in four dimension. In the background of this domain wall solution, we have studied behaviour of the operator dual to certain fermionic mode in the supergravity theory, which does not couple to gravitino or other fermionic modes. In the dual field theory, the domain wall solution corresponds to condensation of a chiral primary operator given by a linear combination of superpotential and  $tr(W_\alpha W^\alpha)$, while the fermionic operator dual to the supergravity mode belongs to a multiplet given by $tr (e^V \bar{W}_{\dot{\alpha}} e^{-V} W^2)$. 

We have artificially dialled the charges and explored existence of normal modes in the space-like region. We found for the charge $q=1$ that follows from supergravity, there is no normal mode. Higher charge $q=4.5$ admits normal modes but at $\omega >0$ leading to gapped spectrum. If we increase charge further, there are normal modes at $\omega=0$ as well. We obtain a dispersion relation for the normal modes. In the time-like region,  for $q=4.5$ and $q=10$ we find peaks of spectral function. The dispersion relation in the time-like region turns out to be quadratic in $k$, considering both the regions a hyperbolic fit matches well. In the case of $q=1$, the charge following from supergravity theory, however, we have not observed any peak.  

Fermionic quasi particles in presence of condensate at zero temperature has a similar gapped spectrum \cite{Chen:2009pt}. Gapped spectra were also found in four dimensional gauged supergravity dual to ABJM model with broken $U(1)$ symmetry \cite{DeWolfe:2015kma,DeWolfe:2016rxk}, where the gap has been attributed to the low charge or particle hole interaction. In the present analysis, it seems that the small charge is responsible for the gapped spectrum. The condensed bosonic operator may play substantial role in the behavior of spectrum of 
fermionic operators considered here. An understanding of the role of the condensed scalar operator in determining the spectrum of the fermionic one, from the perspective of field theory would be interesting. 
 
The five dimensional supergravity obtained after suitable truncation gives rise to several decoupled sectors of fermionic modes. In the present work we have restricted ourself to the case of the fermionic sector consisting of a single fermion. It may be interesting to extend this analysis to the fermionic modes in the other sectors as well. However, those fermions are coupled with one another and also coupled to gravitino and so it calls for a more involved analysis. In the present discussion we have neglected back reaction of the fermions and a natural extension would be to consider it.

%
			\section*{Acknowledgement}
			The work of NR is supported by University Grants Commission of India (UGC India).

\end{document}